\documentclass[prl, twocolumn]{revtex4-2}
\usepackage{amssymb}
\usepackage{amsmath}
\usepackage{epsfig}
\usepackage{bm}
\usepackage{xcolor}

\begin{document}

\title{ Kinetic coefficients of two-dimensional electrons with strong Zeeman splitting  }

\author{ Yu. O. Alekseev, P. S. Alekseev, and A. P. Dmitriev}

\affiliation{  Ioffe  Institute,  194021  St.~Petersburg, Russia }

\begin{abstract}

In modern nanostructures with  very low defect densities,  has recently been realized
a hydrodynamic regime of electric transport, in which two-dimensional (2D)
 electrons form a viscous fluid due to frequent electron-electron collisions.
Many bright transport  phenomena have been observed in these systems.
Of particular interest are two-component hydrodynamic  electron systems,
  where a richer variety of phenomena becomes possible,  than in one-component systems.
A simplest way to implement and control a two-component 2D electron system is to place
 a  structure with 2D electrons in a magnetic field with a large component  in the 2D plane,
  that leads to a Zeeman splitting of the electron energy spectrum into two subbands.
 Here we develop a microscopic model of hydrodynamic transport in such system.
By solving the kinetic equation, we calculate the electron-electron relaxation rates of
  the first and second angular harmonics of the two-component distribution function.
 Then we derive the hydrodynamic balance equations with  the kinetic coefficient containing
these rates.   Namely, are taken into account  the shear viscosity in each fluid component
  and the effect of the  friction between the two components. The last leads to equalization
 of the hydrodynamic velocities in the two subbands. The obtained equations can be used
to explain the results  of puzzling magnetotransport  experiments in ultra-pure nanostructures
in a strong oblique  magnetic field.

\end{abstract}

\maketitle

\section{ 1. Introduction}
Transport properties of ultra-pure 2D electron  systems have been actively studied
since the 1980-s in many directions.  Discovery of the hydrodynamic
regime of electric  transport was an important event in this field.
   Such regime was proposed in the 1960-s by R. N. Gurzhi and co-authors
    for pure bulk metals with the strong electron-phonon
interaction~\cite{Gurzhi_rev,gurzhi1963minimum,gurzhi1995electron},
 and reliably detected in recent ten years in ultra-pure graphene samples,
layered metals~PdCoO$_2$, and GaAs/AlGaAs quantum
wells~\cite{Makkenzi_et_al_2016,Geim_et_al_2016,Levitov_et_al_2016,
alekseev2016negative,Gusev_et_al_AIP_2018,Du_et_al_2022,gusev2018viscous}.
 In 2D systems the hydrodynamic regime  is formed    in a certain temperature range
and  is associated with frequent electron-electron collisions, leading
to the formation of a viscous electron fluid~\cite{f}.

 Different geometries of 2D electron flows were considered: regular and irregular.
As an example of a hydrodynamic effect in an irregular flow geometry,
 whirlpools of 2D electron fluid were realized in a complex-shaped graphene sample
 with several contacts~\cite{Geim_et_al_2016}.
 A simplest type of regular geometry is the Poiseuille flow of
the electron fluid, which is considered in many
 theoretical and experimental works (see, for example,
 Refs.~\cite{alekseev2016negative,Gusev_et_al_AIP_2018}).
Another type of regular geometry is the flow of 2D electrons
through an array of macroscopic localized defects~\cite{d1,d1_new,d2,d2_new}.
 Such defects can be made artificially or  can appear via the
 growth process. The theory of magnetotransport in them was developed in
Refs.~\cite{AP_PS,IV_DG}.

Ohmic and hydrodynamic magnetotransport
in two-component 2D systems has also been actively studied in the last decade.
Two-component electron-hole systems can be realized, for example,
on base of  graphene and other conductors with zero or very small band gap
 by populating  carriers in two closely
 located bands at sufficiently high temperatures.
 Magnetotransport phenomena in graphene are of particular interest
when the chemical potential of the electron-hole system corresponds to
the charge neutrality point, that is,    to equal equilibrium densities of electrons
and holes. In this case, when current flows in a magnetic field,
  electron-hole symmetry is not broken and, consequently, the Hall effect is absent.
In works~\cite{we1,we6,we4,we3}  a theory of classical magnetotransport in
two-component electron-hole systems at the charge neutrality point and away from it
  was constructed for the Ohmic and  hydrodynamic regimes. In these works,
  in particular, it was shown that the formation in samples of  spatially
inhomogeneous flows with high current density in regions near the edges
can lead to strong positive non-saturating linear magnetoresistance.
Such magnetoresistance, possibly, was observed   in experiment~\cite{we2}
and in other experiments.

Of significant interest is also hydrodynamic magnetotransport
in two-component  electron systems, in which there are electrons
 of two types with substantially different  parameters:
  Fermi surface sizes, relaxation times.
 Magnetotransport in such two-component electron system,
 which will be considered
 in this paper, differs significantly
 from the magnetotransport in one-component systems.
Indeed,  inter-electron collisions in a two-component system lead, in addition
to the relaxation of shear stress and the shear viscosity effect
 (which is the case in  one-component systems), also lead to the  friction between
 the two fluid  components, that is,  to the relaxation
of the difference of the hydrodynamic velocities  of these two subsystems.
It is noteworthy that two-component electron-electron  systems also differs
significantly  from  two-component   electron-hole  systems,
 primarily,  due to the sign of charge of the hole component.
  The last  leads to specific effects in transport phenomena, for example,
as it was mentioned above,   at equal electron and hole concentrations
 the Hall field is absent and magnetoresistance can be linear and non-saturating.
  In the two-component electron-electron  system similar effects
  are obviously impossible.

The two-component 2D electron-electron hydrodynamic systems can be
implemented  experimentally, for example, by the two following ways.
First,  such systems are realized in quantum  wells
by populating the two 2D bands  formed
 by the two  lowest quantum confinement levels due to
a sufficiently high doping level and/or temperature~\cite{levin2024bulk}.
Second, a two-component  electron system
can be formed by applying a magnetic field with a large component
in the quantum well plane to a structure with
only one lowest 2D band  occupied  by electrons.
 A strong enough  in-plane magnetic field induces the transformation
of  the  single, degenerated by spin, band
into the two Zeeman subbands~\cite{dai2011response,Hatke_Zudov,small}.

A phenomenological theory of hydrodynamic magnetotransport in a two-component
2D electron system was developed   in  work~\cite{alekseev2023viscous}.
Its predictions are in qualitative agreement
with the results of   experiments~\cite{dai2011response,levin2024bulk},
in which  single-component electron systems  were reconstructed by the described above  ways
into  two-component systems  and  an evolution of the giant negative magnetoresistance
 into the positive saturating magnetoresistance was observed.
 However, in experiment~\cite{levin2024bulk}, the amplitude of  the resulting
positive magnetoresistance turns out to be significantly less than unity, which
differs from the predictions of theory~\cite{alekseev2023viscous}, in which
  saturating magnetoresistance with an amplitude of the order of unity   was predicted
  for a significant difference in parameters of the two components.
 Note also that in theory~\cite{alekseev2023viscous}
the  friction between the two fluid  components due to the electron-electron
scattering was not accounted.

Thus, is relevant a detailed  microscopic theory of magnetotransport
in two-component electron fluids for a quantitative explanation of
experiments~\cite{dai2011response,Hatke_Zudov,levin2024bulk}.
This theory   is to be based on exact calculations of  the relaxation rates
of  different angular  harmonics of the distribution  functions  of
the two fluid components and construction   of the corresponding Navier-Stokes-like
flow equations.  It  should significantly clarify and correct
the phenomenological theory of Ref.~\cite{alekseev2023viscous}. In particular,
the microscopic theory  should quantitatively explain the evolution, with the increase
of the Zeeman splitting,  of the strong negative magnetoresistance,
being characteristic for one-component  viscous electron fluids,
into a positive saturating
magnetoresistance,  observed for two-component systems in experiments
~\cite{dai2011response,levin2024bulk}  (and, possibly, in~\cite{Hatke_Zudov}).

In this work,  we develop such theory for the case of a two-component electron
system  formed  due to a Zeeman splitting of the 2D band
 formed on the lowest size quantization level in a quantum well.
 We  examine  the electron kinetics in  such two-component system
at a sufficiently strong splitting and low temperatures,  when the two
resulting subbands have strongly  different densities
[see Fig.~1($a$)]. We formulate  the definitions of the all  relaxation rates
of the first and second angular  harmonics of the two-component distribution function,
those control the formation of electron flows.  Using the calculated relaxation rates,
we  derive the kinetic coefficients and the relaxation terms
 in the Navier-Stokes-type balance transport equations. Their solution should lead
 to the description  of various magnetotransport effects in  this system,
 in particular, to clarify the mechanism  of the evolution of the magnetoresistance
(relative to the  perpendicular component  of magnetic field)  with the increase
 of the in-plane  magnetic field component.

\section{ 2. Basic equations }
We consider  2D electrons in a quantum well at the first level
of size quantization, to which a magnetic field~$\mathbf{B}$
 with a large in-plane component $\mathbf{B}_{||}$, $B_{||}  \approx B$,
 is applied. As a result, the  Zeeman splitting of this level leads
  to the formation of the two-component electron system consisting
  of the spin-up and spin-down 2D subbands  [see Fig.~1($a$)].
 We consider that the magnitude of  the Zeeman  splitting greatly
exceeds the temperature. It is assumed that impurity concentration
 is very low, therefore,
the scattering of electrons by impurities is not taken into account.
 Below we will denote the quantity $B_{||}$  just as~$B$.

\begin{figure}[t!]
\centerline{\includegraphics[width=1.02 \linewidth]{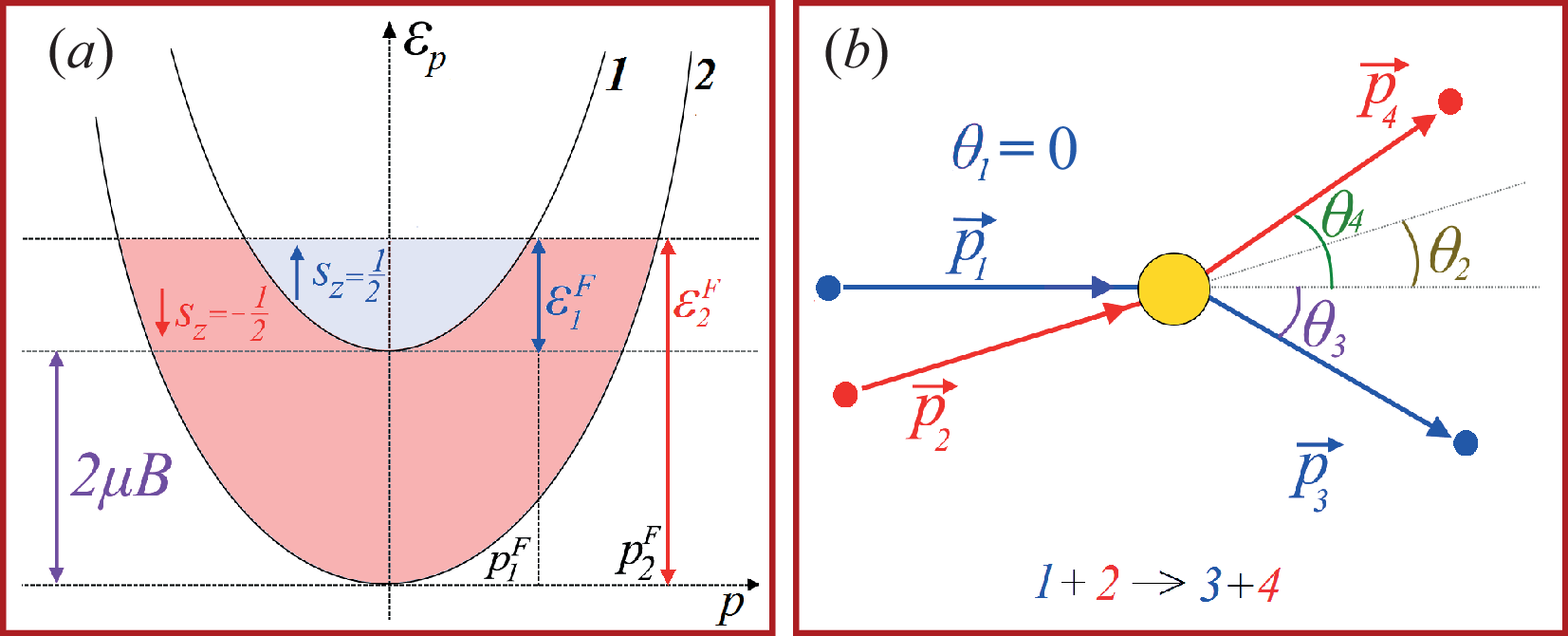}}
\caption{($a$): Energy spectra and equilibrium electron distributions
        of the two components (spin-up and spin-down) of the 2D electron fluid
         in a magnetic field~$\mathbf{B}$
         with a large in-plane  component~$\mathbf{B}_{||}$
        and the corresponding parameters of the two components:
        the Zeeman energy~$\mu B \approx \mu B_{||}$,
        the Fermi energies $\varepsilon^F_{1,2}$,
        and the Fermi momenta~$p^F_{1,2}$.
        Temperature~$T$ is supposed to be low
        as compared with~$\mu B \approx \mu B_{||}$
        and~$\varepsilon^F_{1,2}$.
    ($b$):  Notation for the initial and final momenta
     and scattering angles of two electrons.
    }
\label{parabola}
\end{figure}

Such  system one of the simplest models of two-component electron systems.
The energy spectra of 2D electrons in the Zeeman subbands ``1''~($s_z = 1/2$)
and~``2''~($s_z = - 1/2$) have the form  [see Fig.~\ref{parabola}($a$)]:
\begin{equation}
\varepsilon
_ {1, 2; \, p} = \frac{ p^2}{2m} \pm \mu B  \: ,
\, ,
\end{equation}
where $\mu$ is the electron magnetic momentum and we imply that the Zeeman splitting
 is far larger that temperature, $\mu B\gg T$ (herewith we consider that also
 $\varepsilon _ {1, 2; \, p} \gg T$). Total particle
concentration~$n$ consists of the particle concentrations
in these subbands,~$n_1$ and~$n_2$:
\begin{equation}
    n=n_1+n_2 \, ,~~~
    n_2=\frac{m\varepsilon_2^F}{2\pi\hbar^2}
    \, ,~~~n_1=\frac{m\varepsilon_1^F}{2\pi\hbar^2}
    \,,
\end{equation}
where $\varepsilon_1^F$ and $~\varepsilon_2^F$ are the Fermi energies
in the subbands:
\begin{equation}
   \varepsilon_{1,2}^F=\frac{\pi\hbar^2n}{m}\mp \mu B
   \,.
\end{equation}

For typical GaAs quantum well systems a simple estimates shows that
 the value $\mu B$ can  be of the order of 1~meV at $B \approx B_{||}$=20~T, and
the value $\varepsilon_F$  at~$B=0$
can  be of the order of 3~meV at $n=10^{11}$~cm$^{-2}$.

The system of kinetic equations for the distribution functions
 of the components of the electron system
in a spatially homogeneous state
in the absence of external fields has the form:
\begin{equation}\label{boltzmann}
   \begin{cases}
   \displaystyle
         \partial  f/  \partial t
       \, = \,
       \mathrm{St}_{11}[ \, f(\vec p_1)]+
          \mathrm{St}_{12}[ f(\vec p_1), g(\vec p_2)\,]
       \,,
       \\
       \displaystyle
       \partial g /  \partial  t
       \, = \,
       \mathrm{St}_{22}[\, g(\vec p_2)]+
         \mathrm{St}_{21}[ g(\vec p_2), f(\vec p_1)\,]
       \,,
   \end{cases}
\end{equation}
where  $f = f(\vec p_1,t)$ and $ g = g(\vec p_2,t )$ are
the electron distribution functions of the first
and the second components; $\vec p_1,~\vec p_2$, are the momenta of the electrons
from these components;
$\mathrm{St}_{11}[ f(\vec p_1)]$
and $\mathrm{St}_{22}[ g(\vec p_2)]$ are collision integrals
corresponding to the interaction of particles with identical spin projections
(the intra-band collisions), while
$\mathrm{St}_{12}[ f(\vec p_1), g(\vec p_2)]$
and $\mathrm{St}_{21}[ g(\vec p_2), f(\vec p_1)]$
 are the collision integrals  of particles with opposite projections of spins
 (the inter-band collisions).

The purpose of this work is to find the kinetic coefficients
  of the system. A method of calculation  of the kinetic coefficients
of a single component electron fluid  was described in
Refs.~\cite{pitaevskii2012physical,sykes1970transport} for three-dimensional electron
systems and in Ref.~\cite{Polini,alekseev2020viscosity}  for 2D electron systems.
Kinetic coefficients are expressed in terms   of the relaxation rates
of the inequilibrium   parts  of the distribution function.   In linear approximation
by small inequilibrium  corrections  to the equilibrium distribution functions,
we describe the electron distributions in the two subbands by  the following ansatz:
\begin{eqnarray}
 \label{anz}
 \displaystyle
     f(\vec p_1,t)\approx f^F_1(\varepsilon_1)-\varepsilon_1
     \,
     (f^F_1(\varepsilon_1))'\sum_{m=1}^\infty\Phi_1^{(m)}( t)\,e^{im\theta_1}
     ,
     \label{dec_f}
     \nonumber \\
     g(\vec p_2,t)\approx f^F_2(\varepsilon_2)-\varepsilon_2
     \,
     (f^F_2(\varepsilon_2))'\sum_{m=1}^\infty\Phi_2^{(m)}(t)\,e^{im\theta_2}
     ,
\end{eqnarray}
where $\varepsilon_{1,2} = p_{1,2}^2/(2m) $
 are the energies of electrons from the two components;
$f^F_{1,2} (\varepsilon_{1,2})
 =(1+e^{[\varepsilon_{1,2}-\varepsilon_{1,2}^F]/T})^{-1}$ are
their equilibrium distribution functions;
$(f_{1,2}^F)'=\partial f^F_{1,2}/\partial\varepsilon_{1,2} $;
$\theta_i$ is the angle of the momentum direction $\vec{p}$ of the particle
in the  $i$-th component of the system, $i=1,2$ [see Fig.~1($b$)].
 Decompositions~(\ref{anz}), in which the  inequilibrium  parts
 of the electron distributions are
the products of the derivatives of the equilibrium  parts, $ (f^F_{1,2})' $,
 and the factors depending only   on the  angles~$\theta_i$, implie that:
  (i)~the relaxation by energy  of inequilibrium   perturbations
is much faster that the relaxation by the angles the momentum directions;
  (ii)~the Zeeman splitting~$\varepsilon^F_2 - \varepsilon^F_1 =\mu B $
is much greater than temperature~$T$.

It is known that, in the linear approximation by the inequilibrium part
of the distribution functions,
  different angular harmonics~$\Phi_i^{(m')}( t) \, e^{im'\theta_{i}}$ and
  ~$\Phi_i^{(m'')}( t) \, e^{im''\theta_{i}}$, defined by Eqs.~(\ref{anz}),
  evolve independently.

Two subband components, $i = 1 , 2 $, of
a given $m$-th angular harmonic $\Phi_i^{(m)}( t) \, e^{im\theta_{i}}$
of the two-subband distribution function
do not relax independently in the general case.
That is, their evolution is described by the equation:
\begin{eqnarray}
        \dot{\vec \Phi}^{(m)}=-\Gamma^{(m)}\vec \Phi^{(m)}=-\begin{pmatrix}
        \gamma_{11}^{(m)}&\gamma_{12}^{(m)}
        \\\\
        \gamma_{21}^{(m)}&\gamma_{22}^{(m)}
    \end{pmatrix}\vec \Phi^{(m)}
    \: ,
    \label{gammadefinition}
\end{eqnarray}
where    $\Gamma^{(m)}$ is the relaxation matrix  of the $m$-th harmonic and
the vector $ \vec\Phi^{(m)}$ consists of the $m$th harmonic of
the distribution functions in each of the components:
\begin{eqnarray}
    \vec\Phi^{(m)} (t) = \begin{pmatrix}
        \Phi_1^{(m)} (t)
        \\\\
        \Phi_2^{(m)}(t)
    \end{pmatrix}
    .
\end{eqnarray}
Note right away that when the sign of the magnetic field
component~$\mathbf{B}_{||}$ changes, the subbands~``1'' and~``2'' change
one to another by their energy positions
and, correspondingly,
    the elements of the matrices $\Gamma^{(m)}$ exchange their places:
$\gamma_{11}^{(m)}\longleftrightarrow\gamma_{22}^{(m)}$,
 $\gamma_{12}^{(m)}\longleftrightarrow\gamma_{21}^{(m)}$.
This property of  relaxation rates
will be further used to check their direct  calculations:
 instead of calculating both~$\gamma_{11}^{(m)}$
 and~$\gamma_{22}^{(m)}$  (and~$\gamma_{12}^{(m)}, \gamma_{21}^{(m)}$) at
the positive Zeeman splitting,~$B_{||} >0$,
we can find only  $\gamma_{11}^{(m)}$ (and~$\gamma_{12}^{(m)}$)
for both positive and negative~$B_{||} $.

The linearized collision integral for the electro-electron collisions
 within the first
component~``1'' for the $m$-harmonic term in decomposition~(\ref{anz})
 of~$f(\vec{p}_1,t)$ has the form:
\begin{eqnarray}
 \label{St_11}
    \mathrm{St}_{11} \big[\, \Phi_1 ^{(m)} (t) \, e^{im\theta} \, \big]
    (\varepsilon_1, \theta_1) =
    \nonumber\\
    -\frac{m^2}{(2\pi\hbar)^4T}\frac{2\pi }{\hbar}
     \int |M''|^2  f_1^Ff_2^F(1-f_3^F)(1-f_4^F)
    \times\nonumber\\
      \times\delta[\varepsilon_1+\varepsilon_2-\varepsilon_3
      -(\vec p_1+\vec p_2-\vec p_3)^2/2m]
      \times\nonumber\\
      \times
     \big[ \,  \varepsilon_1\Phi_1^{(m)} (t)
     (e^{im\theta_1}-e^{im\theta_3+im\theta_1}) +
      \varepsilon_1\Phi_1^{(m)}(t)
      \times
      \nonumber\\
      \times
      (e^{im\theta_2+im\theta_1}-e^{im\theta_4+im\theta_1}) \, \big]
      d\varepsilon_2d\varepsilon_3d\theta_2d\theta_3
      ,
\end{eqnarray}
and similarly for   $\mathrm{St}_{22} [\, \Phi_2 ^{(m)} (t) \, e^{im\theta} \, ]$.
 Here the squared matrix element $|M''|^2$ accounts the proper antisymmetrized
initial and final wave functions   of the scattering electrons  with
the space and spin components    (presented in the next section).
The following notations are used hereafter:
\begin{equation}
f_i^F = f^F (\varepsilon _i)
\, ,
\;\;\;
f_i = f(\vec p_i, t)
\, ,
\;\; \;
g_i = g(\vec p_i ,t)
\, .
\end{equation}

The linearized collision integral for the inter-band collisions
involving electrons
from both the first and second components is as follows:
\begin{eqnarray}
\label{St_12}
    \mathrm{St}_{12}\big[\, \Phi_1 ^{(m)} (t)  \, e^{im\theta}
    \, , \,
    \Phi_2^{(m)} (t) \, e^{im\theta} \,  \big] (\varepsilon_1, \theta_1)
    =
        \nonumber\\
    -\frac{m^2}{(2\pi\hbar)^4T}\frac{2\pi }{\hbar}
    \int |M'|^2 f_1^Fg_2^F(1-f_3^F)(1-g_4^F)\times\nonumber\\
      \times\delta[\varepsilon_1+\varepsilon_2-\varepsilon_3
      -(\vec p_1+\vec p_2-\vec p_3)^2/2m]
      \times\nonumber\\
      \times
      \big[ \, \varepsilon_1\Phi^{(m)} _1(t)(e^{im\theta_1}
      -e^{im\theta_3+im\theta_1})
      +
      \varepsilon_2\Phi^{(m)}_2(t)
      \times
      \nonumber\\
      \times
      (e^{im\theta_2+im\theta_1}-e^{im\theta_4+im\theta_1}) \, \big]
      d\varepsilon_2d\varepsilon_3d\theta_2d\theta_3
      .
\end{eqnarray}
Here the squared matrix element $|M'|^2$ accounts all the proper antisymmetrizes
   initial and final wave functions
    of the scattering electrons with the space and spin components (see the next section).

\section{ 3. Matrix element of electron-electron scattering }
We denote by $\vec P$ and $\vec Q$ the  transmitted momentums
in the direct and exchange contribution in the intra-subband and inter-subband
scattering matrix elements  $M''$ and $M'$ :
\begin{eqnarray}
    \vec p_1-\vec p_3 = - \vec{p}_2 + \vec{p}_4 = \vec P\,,
    \nonumber \\
    \vec p_1-\vec p_4 = - \vec{p}_2 + \vec{p}_3= \vec Q
    \,,
\end{eqnarray}
Let us choose the Rytova-Keldysh potential~\cite{rytova2018screened}
  as a realistic law of the interaction
of degenerated 2D electron in GaAs quantum wells:
\begin{eqnarray}
 \label{U}
    U( P)=\frac{2\pi \,( e^2 /\kappa ) \, a_B}{a_B P/\hbar+2}
    \:,
\end{eqnarray}
where $a_B$ is the Bohr radius, $\kappa$ is the dielectric constant
 the environment of a 2D layer.

The wave functions have the form:
\begin{eqnarray}
    \boldsymbol \Psi_{1\uparrow,2\downarrow}
    =
    \frac{  e^{i\vec k_1\vec x_1+i\vec k_2\vec x_2}|\uparrow \downarrow\rangle
    -
    e^{i\vec k_2\vec x_1+i\vec k_1\vec x_2}|\downarrow\uparrow \rangle  }{\sqrt 2}  \label{psi'}
    \,,
    \\
    \boldsymbol \Psi_{3\uparrow,4\downarrow}=\frac{e^{i\vec k_3\vec x_2+i\vec k_4\vec x_1}|
    \uparrow \downarrow\rangle
    -
     e^{i\vec k_4\vec x_2+i\vec k_3\vec x_1}|\downarrow\uparrow \rangle  }{\sqrt 2}
    \, .
    \label{psi'2}
\end{eqnarray}
Here, in the first term of~$\boldsymbol \Psi_{1\uparrow,2\downarrow}$,
  the first component of the spinor~$|\uparrow\rangle$
corresponds to the momentum~$\vec k_1 = \vec p_1/\hbar$,
 the second spinor component~$|\downarrow\rangle$
corresponds to the momentum $\vec k_2= \vec p_2/\hbar$, and similarly in the other terms.
In other words, provided the two scattering electrons were different, by some reason,
 than in the initial state the momentum of a particle
with a positive spin projection is $\vec k_1$, the momentum of a particle
with a negative spin projection is $\vec k_2$, and in the final state
the momentum of a particle with a positive spin projection is $\vec k_3$,
the momentum of a particle with a negative spin projection is $\vec k_4$.
As the two initial and the two final electrons are undistinguished actually,
 therefore these initial and final states need to be antisymmetrized,
 as it is done in expressions~(\ref{psi'}) and~(\ref{psi'2}).

Thus, functions~(\ref{psi'2}) of a pair of particles are characterized
 by projections of spins for each $\vec{k}_i$,
  rather than the total spin, so it cannot
  written as a product of the orbital and spin functions and so
is not a conventional singlet state. Such wave functions
 between which the ee-scattering occurs
  differ sufficiently
 from the ones in the case of zero magnetic field and no spin-orbit interaction,
 when the usual singlet and triplet function, being the products
 of the orbital and spin functions, can be used.

The corresponding  direct and the exchange matrix elements  for transitions
between such ``inter-subband'' states have the form:
\begin{eqnarray}
    M'(P)=
    \langle \, \boldsymbol  \Psi_{3\uparrow,4\downarrow} \,  |\,
    U
     \,  |\,  \boldsymbol \Psi_{1\uparrow,2\downarrow} \, \rangle=U(P)
    \,,
    \nonumber
    \\
    M'(Q)=\langle\, \boldsymbol \Psi_{ 3\downarrow , 4\uparrow }\,|\,
    U
    \,|\,\boldsymbol \Psi_{1\uparrow,2\downarrow}\,\rangle=U(Q)
    \,.
\end{eqnarray}

The intra-subband wave functions have the form:
\begin{eqnarray}
     \boldsymbol\Psi_{1\uparrow,2\uparrow}
     =
     \frac{1}{\sqrt 2}(e^{i\vec k_1\vec x_1+i\vec k_2\vec x_2}
     -
     e^{i\vec k_1\vec x_2+i\vec k_2\vec x_1})|\uparrow\uparrow\rangle
     \,,
     \nonumber
     \\
     \boldsymbol\Psi_{1\downarrow,2\downarrow}
     =
     \frac{1}{\sqrt 2}(e^{i\vec k_1\vec x_1+i\vec k_2\vec x_2}
     -
     e^{i\vec k_1\vec x_2+i\vec k_2\vec x_1})|\downarrow\downarrow\rangle
     \,.
     \label{intra-states}
\end{eqnarray}
These are the triplet-like wave functions (namely,
 the two components of the triplet with $s_z = \pm 1$),
which are products of symmetric spin functions and
antisymmetric orbital functions. The matrix elements of the transition
between intra-subband states~(\ref{intra-states})
have the form:
\begin{equation}
   M''(P,Q)= \langle  \, \boldsymbol\Psi_{3\uparrow,4\uparrow} \, | \, U \, |
   \, \boldsymbol\Psi_{1\uparrow,2\uparrow} \, \rangle =U(P)-U(Q)
   \,.
\end{equation}
The square of this matrix element has the form:
\begin{equation}
\begin{array}{c}
      |M''(P,Q)|^2=|U(P)-U(Q)|^2=
      \\\\
      \displaystyle
      =
     \Big( 2\pi \frac{ e^2 \, a_B } { \kappa } \Big) ^2 \,
      \Big(\frac{ \displaystyle   1 }{ \displaystyle  a_B P /\hbar+2}
      -
      \frac{ \displaystyle  1 }{ \displaystyle  a_B Q/\hbar+2} \Big)^2
      .
      \end{array}
      \label{M''}
\end{equation}
This expression includes the terms~$U(P)^2$ and~$U(Q)^2$ as well as
the term~$U(P)\, U(Q)$, that it is not a function
 of only $Q$ or only $P$.

\section{ 4.  Relaxation rates of angular harmonics }
\subsection{ 4.1. Definitions and starting equations }
In this section, we present the definitions and procedure of calculation
for  the relaxation rates of angular perturbations of the distribution functions
 due to   the scattering of the electrons  in the two subbands.

As it was mentioned above,
 the conditions $T\ll\varepsilon_{1,2}^F$ and $|\varepsilon_1^F-\varepsilon_2^F|\gg T$
are met.   In this case it follows from the energy and momentum conservation laws
 and the Fermi distribution
  of equilibrium electrons and corresponding distribution of empty states
  [described by the factor~$f_1^F f_2^F (1-f_3^F) (1-f_4^F) $
   in Eq.~(\ref{St_11}) and~(\ref{St_12})]
that the initial and final momenta of electrons
in both subband for the inter-band scattering
 as well as intra-band collisions lie near the corresponding Fermi surfaces:
\begin{equation}
\label{near}
 |\vec p_{1,3}|\approx p_{1}^F
 \,, \quad
 |\vec p_{2,4}|\approx p_{2}^F
 \, .
\end{equation}
In this approximation, for inter-band collisions
 the scattering angles and the energy transferred
during scattering become nearly independent [see Fig.~2($a$)].
 For them, significant contributions
to the relaxation rate comes from the scattering processes with collision angles over
the entire range, $\theta_2\sim 1$,  while the scattering by incoming electrons
with the opposite and the small angles, $\theta_2\sim\pi$ and $\theta_2\sim 0 $,
is not highlighted in any way. It is well-known that the contributions from the last diapason of angles
dominate in electron-electron scattering
in one-components systems~(see, for example, Ref.~\cite{alekseev2020viscosity})
 and, analogously,  for the intra-band collision
 in the considered here two-component system [see Fig.~2($b$)]

\begin{figure}[t!]
\centerline{\includegraphics[width=0.99 \linewidth]{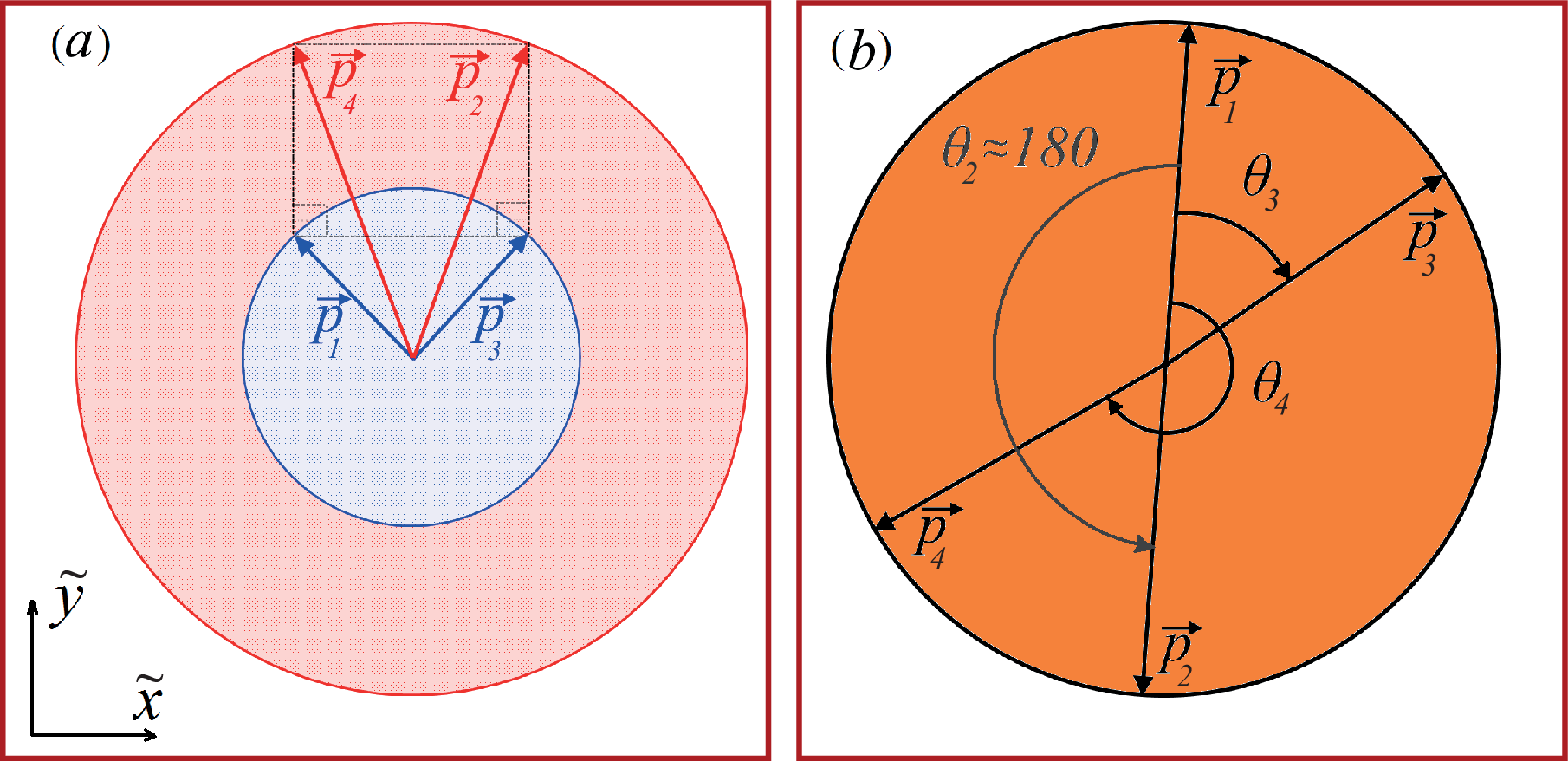}}
      \caption{Kinematics of inter-subband  ($a$) and intra-subband  collisions~($b$).
      The initial and final momenta
       correspond to the conservation of the total momentum
       during collisions and small energy changes.  In panel~($a$)
       the direction~$\tilde{y}$
    corresponds to the initial total momentum of the electron pair,
     $\vec{p}_1+\vec{p}_2$, that is $\vec{p}_{1,\, \widetilde{x}}
     +
     \vec{p}_{2,\, \widetilde{x}} =0 $.
     In intra-subband  collisions, the {\em head-on} scattering on the electrons with almost
     opposite momenta, $ \vec{p}_2 \approx - \vec{p}_1  $, dominates.
     }
      \label{kinematika}
  \end{figure}

Note that the last property of collision kinematics for the inter-subband
scattering  in  two-component systems differs it  sharply
from the case of   one-component Fermi systems, where  the
 {\em head-on} collisions, for which $\theta_2\sim \pi$
is dominant~\cite{gurzhi1995electron,alekseev2020viscosity}.
Herewith, the intra-subband scattering  in  two-component systems,
obviously, have the same properties as ee-collisions in
one-component systems.

Let us denote by $\alpha^{(m)}$ and $\beta^{(m)}$ the contributions of inter-subband
and intra-subband collisions to the overall relaxation rates:
\begin{eqnarray}
\Gamma^{(m)}={\alpha^{(m)}}+{\beta^{(m)}}
\, ,
\label{sum}
\end{eqnarray}
where
\begin{eqnarray}
    \alpha^{(m)}=\begin{pmatrix}
        \alpha_{11}^{(m)}& \alpha_{12}^{(m)}\\
         \alpha_{21}^{(m)}& \alpha_{22}^{(m)}
    \end{pmatrix},~~~\beta^{(m)}=\begin{pmatrix}
        \beta_{11}^{(m)}& \beta_{12}^{(m)}\\
         \beta_{21}^{(m)}& \beta_{22}^{(m)}
    \end{pmatrix}
    .
\end{eqnarray}
These values are calculated based on the obtained matrix elements,
 taking into account the assumptions made [in particular, the simplified form
  of the perturbed parts of the distribution functions in Eqs.~(\ref{anz})].

First, let us describe the calculation of the contribution
of inter-subband collisions into the scattering rates,~${\alpha^{(m)}}$.

The equation for the amplitudes  $\Phi_1^{(m)}$ of the $m$-the harmonics,
following from the inter-band collision integrals~$\mathrm{St}_{12}$~(\ref{St_12}),
takes the form:
\begin{eqnarray}
\varepsilon_1\, (f_1^F)'(\varepsilon_1)\,\dot\Phi_1^{(m)}(t) \, e^{im\theta_1}
=\nonumber\\=
\frac{m^2}{(2\pi\hbar)^4T}\frac{2\pi }{\hbar}\int |M'|^2f_1^Fg_2^F(1-f_3^F)(1-g_4^F)
\times\nonumber\\\times
 \delta\big[  \varepsilon_1+\varepsilon_2-\varepsilon_3
 -
 (\vec p_1+\vec p_2-\vec p_3)^2/2m  \big] \times \nonumber\\
      \times
      \left[ \varepsilon_1\Phi_1^{(m)}(t)(e^{im\theta_1}-e^{im\theta_3+im\theta_1})
      +\varepsilon_2\Phi_2^{(m)}(t)
      \times
      \right.
      \nonumber\\
      \left.
      \times
      (e^{im\theta_2+im\theta_1}-e^{im\theta_4+im\theta_1})\right]
      d\varepsilon_2d\varepsilon_3d\theta_2d\theta_3
      \,.
      \label{St_for_Phi_1_m}
      \end{eqnarray}
For the amplitude $\Phi_2$, the formula turns out to be similar.

Dividing both sides of equation~(\ref{St_for_Phi_1_m}) by  $e^{im\theta_1}$
and integrating by $\varepsilon_1$, we obtain:
\begin{eqnarray}
  \varepsilon_1^F  \,  \dot\Phi_1^{(m)}=\nonumber\\
  -\frac{m^2}{(2\pi\hbar)^4T}\frac{2\pi }{\hbar}\int |M'|^2f_1^Fg_2^F(1-f_3^F)(1-g_4^F)
  \times\nonumber\\ \times \:
  \delta\big[  \varepsilon_1+\varepsilon_2-\varepsilon_3
  -
  (\vec p_1+\vec p_2-\vec p_3)^2/2m  \big]\times\nonumber\\
      \times
      \left[ \varepsilon_1\Phi_1^{(m)}(1-e^{im\theta_3})\right.  +
       \nonumber\\
       +\left.\varepsilon_2\Phi_2^{(m)}(e^{im\theta_2}-e^{im\theta_4}) \right]
       d\varepsilon_1d\varepsilon_2d\varepsilon_3 d\theta_2d\theta_3.
       \label{cat}
\end{eqnarray}
We remind that in limit of low temperatures,
 $T\ll \varepsilon^F_{1,2},\,\varepsilon^F_{1} - \varepsilon^F_{2} $,
it follows from the energy and momentum conservation laws
 and the Fermi statistics, that   the initial and final electron states
 lie near the Fermi surfaces [see  Eq.~(\ref{near})].
  In particular, the relations $  \varepsilon_{1,3}\approx\varepsilon_{1}^F  $,
 $  \varepsilon_{2,4}\approx\varepsilon_{2}^F $ are implied in Eq.~(\ref{cat})
  in the matrix element and in the delta-function. It is  important that, as a result,
in the matrix element $|M'|^2$ the spin subbands,~$\uparrow$ and~$\downarrow$,
and the momentum magnitudes~$p _{1,2}\approx p_{1,2}^F$ of the initial and final states
 correspond one to another, thus only the direct contribution presents in~$|M'|^2$:
\begin{equation}
 |M'|^2  \, \approx \, |M' (P)|^2
 \:,
\end{equation}
with the precision of the order of the small parameter~$T/\varepsilon_{1,2}^F$.
  As a result, we have from Eq.~(\ref{cat}):
\begin{eqnarray}
2 (\varepsilon_1^F)^2 \dot\Phi_1
\,=\, - \,
\frac{m^2}{(2\pi\hbar)^4T}\frac{2\pi }{\hbar}
\times
\nonumber\\
\times
  \int f_1^Fg_2^F(1-f_3^F)(1-g_4^F)
    d\varepsilon_1d\varepsilon_2d\varepsilon_3
   \int|M'|^2
   \times
   \nonumber  \\
        \times \:
     \delta\big[ \cos\theta_3-1- (p_2^F/p_1^F)(\cos\theta_2
      -\cos(\theta_2-\theta_3)) \big]
       \nonumber\\
        \times
        \big[ \varepsilon_1^F \, \Phi_1(1-e^{im\theta_3})  +
    \nonumber \\
    +
   \varepsilon_2^F \, \Phi_2(e^{im\theta_2}-e^{im(\theta_3-\theta_2)})\big]
   \:d\theta_2d\theta_3
  \,.
 \label{rel_from_St__M'}
\end{eqnarray}
After integration by particle energies, we obtain from Eq.~(\ref{rel_from_St__M'}):
\begin{eqnarray}
2 (\varepsilon_1^F)^2 \dot\Phi_1
=
-\frac{2\pi^2T^2}{3}\frac{m^2}{(2\pi\hbar)^4}\frac{2\pi}{\hbar}
 \int|M'|^2
 \times\nonumber
 \\
\times
\delta\big[ \cos\theta_3-1-(p_2^F/p_1^F)(\cos\theta_2-\cos(\theta_2-\theta_3)) \big]
\times\nonumber\\\times \big[ \varepsilon_1^F\Phi_1(1-e^{im\theta_3}) +
\nonumber\\
+
\varepsilon_2^F\Phi_2(e^{im\theta_2}-e^{im(\theta_3-\theta_2)})\big]
 d\theta_2d\theta_3.
 \label{calc1}
\end{eqnarray}
It should be emphasized that when the difference  $\varepsilon_1^F-\varepsilon_2^F$,
becomes comparable with~$T$, there appear singularities in integral~(\ref{calc1}),
and it loses its applicability.

Due to the momentum and energy conserving  laws in scattering,
the angle $\theta_3$ becomes a definite function of the angle $\theta_2$:
$ \theta_3  = \hat\theta_3 (\theta_2) $.
Namely,  the delta-function in the scattering probability in the collision operator
can be expressed,  using the function $\hat\theta_3(\theta_2)$, as:
\begin{eqnarray}
    \delta \big[  \cos\theta_3-1-(\cos\theta_2-\cos(\theta_2-\theta_3))(p_2^F/p_1^F)
     \big]
    =\nonumber\\=
    \frac{\delta(\theta_3-\hat\theta_3(\theta_2) \:)}{(p_2^F/p_1^F)\, |\sin\theta_2|}
    ,
    \label{delta}
\end{eqnarray}
where the scattering  angle~$\hat\theta_3(\theta_2)$ of the electrons
 in the ``1'' subband has the form:
\begin{eqnarray}
    \tan ( \hat\theta_3  /  2  ) =\frac{\sin\theta_2}{p^F_1/p^F_2+\cos\theta_2}
    \:.
    \label{theta3}
\end{eqnarray}
The last dependence is illustrated by  Fig.~\ref{kinematika}($a$).
Combining (\ref{theta3}) and (\ref{calc1}), we obtain the following expressions
for the relaxation rates defined by  Eqs.~(\ref{gammadefinition}) and~(\ref{sum}):
\begin{eqnarray}
    {\alpha_{11}^{(m)}}=\frac{ H }{p_1^Fp_2^F}
    \int\limits_0^{2\pi}|M'|^2\frac{1-e^{im\hat\theta_3[\theta_2]}}{|\sin\theta_2|}d\theta_2
  \label{alfa_11_res}
\end{eqnarray}
and
\begin{eqnarray}
        {\alpha_{12}^{(m)}}=\frac{H \, p_2^F}{(p_1^F)^3}
        \int\limits_0^{2\pi}|M'|^2\frac{e^{im\theta_2}
        -
        e^{im(\hat\theta_3[\theta_2]-\theta_2)}}{|\sin\theta_2|}d\theta_2
        \, ,
        \label{alfa_12_res}
\end{eqnarray}
where $H = [1/(12\pi)](m^3 T^2 / \hbar^5) $ is the temperature-dependent factor.

From the equations for the time derivative of the distribution function
amplitude~$\dot{\Phi}_2$, analogous
to Eqs.(\ref{St_for_Phi_1_m})-(\ref{calc1}), and, equivalently,
from Eqs.~(\ref{alfa_11_res}) and~(\ref{alfa_12_res})
by swapping  the indices ``1'' and ``2'', we obtain :
\begin{eqnarray}
    {\alpha_{22}^{(m)}}= \frac{H}{p_1^Fp_2^F}
    \int\limits_0^{2\pi}|M'|^2
    \frac{1-e^{im\hat{\hat\theta}_3[\theta_2]}}{|\sin\theta_2|}d\theta_2
\end{eqnarray}
and
\begin{eqnarray}
        {\alpha_{21}^{(m)}}=\frac{H\,p_1^F}{(p_2^F)^3}
        \int\limits_0^{2\pi}|M'|^2\frac{e^{im\theta_2}
        -
        e^{im(\hat{\hat\theta}_3[\theta_2]-\theta_2)}}
                               {|\sin\theta_2|}d\theta_2
        \:,
\end{eqnarray}
where  the scattering  angle~$\hat\theta_3(\theta_2)$ of the electrons
 in the ``2'' subband  $\hat{\hat\theta}_3 (\theta_2)$ is:
\begin{eqnarray}
 \tan (  \hat{\hat\theta}_3  /  2 )
=\frac{\sin\theta_2}{p^F_2/p^F_1+\cos\theta_2}
\:.
\end{eqnarray}

Second, let us perform similar calculations for
the intra-subband scattering rates~$\beta ^{(m)}$.
 We start from the intra-band collision integrals~$\mathrm{St}_{11,22}$~(\ref{St_11})
 and do the similar calculations as we have performed above for the inter-band scattering.

First of all, due to the momentum conservation law,  the intra-subband scattering does not
 affect the relaxation of the $m=1$ harmonic controlling the relaxation of
 the electron flows. That is, we just have: $\beta^{(1)} =0 $.

As we noted above, the ``head-on'' collisions
in each subbands of a given electron with the electrons
having the  angles~$\theta_2\approx\pi$ provide the main contribution.
Integration is performed using $\theta_3$, so one need to return to the left side
of formula~(\ref{delta}), put~$p_1^F=p_2^F$,  and choose a solution corresponding
to nontrivial scattering $\theta_3 \neq 0$.  As a results, the delta-function in
the initial variables in the collision integral for the subband~``1'' takes the form:
\begin{eqnarray}
    \delta\Big[\varepsilon_1+\varepsilon_2-\varepsilon_3
    -
    \frac{(\vec p_1+\vec p_2-\vec p_3)^2}{2m}\Big]
    \approx
    \frac{\delta(\theta_2-\pi)}{\varepsilon_1^F|\sin\theta_3|}
    \:.
     \label{delta_for_beta}
\end{eqnarray}

It is important that  the intra-band collisions,
both the direct and exchange contributions
 do enter in the matrix element $|M''|^2$~(\ref{M''}),
 similarly to the case of one component systems
 (see Ref.~\cite{alekseev2020viscosity}).

Using~(\ref{delta_for_beta}), we obtain for the contributions
of the intra-subband collisions  to the relaxation rates of the $m$-th harmonic
of the distribution function:
\begin{eqnarray}
 \label{res_beta_m_11}
    {\beta_{11}^{(m)}}
    =
    (1+(-1)^m)\frac{H}{(p_1^F)^2}
    \int\limits_0^{2\pi}|M''|^2
    \frac{1-e^{im\theta_3}}{|\sin\theta_3|}d\theta_3
\end{eqnarray}
and
\begin{eqnarray}
 \label{res_beta_m_22}
   {\beta_{22}^{(m)}}
   =
   (1+(-1)^m \frac{H}{(p_2^F)^2}
   \int\limits_0^{2\pi}|M''|^2
   \frac{1-e^{im\theta_3}}{|\sin\theta_3|}d\theta_3
   \:.
   \label{inner}
\end{eqnarray}

It is important to note that in our approximation of small~$T$, the contributions
of intra-subband collisions to the relaxation rates of odd harmonics are equal to zero.
Note that  rigorous calculations for inter-particle collisions
in a single-component system yields
parametrically low relaxation rates of the odd rates~\cite{alekseev2020viscosity},
for example:~$\gamma _{11,22} ^{(3)}
 \sim \gamma _{11,22} ^{(2)} \, ( T/\varepsilon ^F_{1,2})^2$.

\subsection{ 4.2. Explicit form of relaxation rates and their properties }
In this subsection we present the expressions for the relaxation rates~$\alpha^{(m)}$
and~$\beta^{(m)}$ in an explicit form.

Let us denote:
\begin{eqnarray}
\cos\theta_2=x \, ,~~~a=p_1^F \,/ \,  p_2^F \, .
\label{X}
\end{eqnarray}
    The transmitted momentum in inter-subband collisions will take the form:
\begin{eqnarray}
    P=2p_1^F\sin\theta_3/2=2p_1^F\sqrt\frac{1-\cos^2\theta_2}{1+a^2+2a\cos\theta_2}
    =
    \nonumber\\
    =
    2p_1^F\sqrt\frac{1-x^2}{a^2+2ax+1}
    \:.
\end{eqnarray}
For direct matrix element of the inter-subband process, we have:
\begin{eqnarray}
\label{35}
    |M'|^2 = \frac{4\pi ^2 (e^2/\kappa)^2 a_B^2}{(a_BP/\hbar+2)^2}
    =
      \nonumber\\
    =
    \frac{ \displaystyle 4\pi ^2 (e^2 a_B /\kappa)^2}
    {\displaystyle  \left(\frac{a_Bp_1^F}{\hbar}
     \sqrt\frac{1-x^2}{a^2+2ax+1}+2\right)^2}
    \:.
\end{eqnarray}
By expressing $\theta_3$ in terms of $\theta_2$ and substituting $x=\cos\theta_2$,
below we will  transform all the above integrals in Eqs.~(\ref{alfa_11_res})-(\ref{inner})
to a more convenient form.

Note that taking the limit of $a\rightarrow 1$ in the final expressions
for the rates is incorrect, even if at $a=1$ they turn out to be finite.
Indeed, the integrands in Eqs.~(\ref{alfa_11_res})-(\ref{inner})
are not uniformly continuous functions of $a$ and $x$,
in the vicinity of $a=1,~~x=-1$, so one cannot take just the limit of $a\rightarrow 1$
after calculating the integrals. Nevertheless, the analysis of the
behavior of the rates calculated within the described approximation in the limits
$a\rightarrow 1 \pm 0$, that correspond to~$p_1 ^F >p_2^F$ and~$p_1^F<p_2^F$
will be of interest.

The relaxation rate of the first harmonic takes the form
[here we use Eqs.~(\ref{X})-(\ref{35})]:
\begin{eqnarray}
\label{alf_res__11}
    {\alpha_{11}^{(1)}}
    = \frac{H}{p_1^F p_2^F } \int_0^{2\pi}
    \frac{1-\cos\hat\theta_3}{|\sin\theta_2|}|M'|^2 d\theta_2
    = \nonumber \\
    = C\int\limits_{-1}^1
    \frac{dx}{
    \displaystyle
    \left(\frac{\sqrt{1-x^2}  }{\sqrt{2} r_s}+\sqrt{a^2+2ax+1}\right)^2}
    \:,
\end{eqnarray}
where
\begin{eqnarray}
\displaystyle
     C=  4\pi ^2  \Big( \frac{e^2 a_B}{\kappa }\Big)^2 \frac{H }{p_1^F p_2^F  }
     = \frac{ \pi}{3}
     \displaystyle  \frac{m\,T^2}{ \displaystyle \hbar \,  p_1^Fp_2^F}
     \label{c}
\end{eqnarray}
and $r_s = r_{s,1} = 1/(a_B \sqrt{\pi n_1})$ is the parameter characterizing
 the relative energy of the electron-electron interaction in the subband~``1''.
At any~$a$ and~$r_s$ this integral
cannot be calculated analytically, however this becomes
possible at limiting values of this parameters
(such results will be presented in the next section).

There is no need in direct calculations, the remaining elements
of the matrix~$\alpha^{(1)}$ of the inter-subband relaxation rates of
the first moment, ${\alpha_{12}^{(1)}},~{\alpha_{12}^{(1)}}$, and~$\alpha_{22}^{(1)}$.
Indeed, these values can be expressed in terms of $\alpha_{11}^{(1)}$.
First, $\alpha_{22}^{(1)}$ is obtained simply by rearranging the indices,
 and, second, the off-diagonal components $\alpha_{12,21}^{(1)}$ can be obtained
from the momentum conservation law and the definition
of~$\alpha^{(1)}$ via Eqs.~(\ref{gammadefinition}) and~(\ref{sum}).
The result for all components of the matrix $\alpha^{(1)}$ is:
\begin{eqnarray}
    \alpha^{(1)}={\alpha_{11}^{(1)}}\begin{pmatrix}
        \displaystyle
        1 &  \displaystyle - p_2^F/ p_1^F
        \\\\
        \displaystyle
        - (p_1^F / p_2^F)^3   &
        \displaystyle  (p_1^F  / p_2^F)^2
    \end{pmatrix}
    =
    \nonumber \\
        \nonumber \\
    =\alpha_{11}^{(1)}\begin{pmatrix}
        1 & -1/a
        \\
        -a^3 & a^2
    \end{pmatrix}
    \,.
    \label{alf_res}
\end{eqnarray}

An analysis shows that one of the eigenvectors of this matrix corresponds to a flow with
the same velocities in the two subbands, and the other with zero total current.
 Below we describe this result in detail.

First, the following relation between the components
of the distribution functions  for two the sates of the two fluid components~``1'' and~``2''
  with a common  mean velocity $V=V_1=V_2$ takes place:
\begin{eqnarray}
   \Phi^{(1)}_1p_1^F = \Phi^{(1)}_2p_2^F
\,.
\label{relat}
\end{eqnarray}
This comes  from the well-known form of the perturbation to the distribution function,
corresponding to a non-zero  flow:~$ \delta f_i (\vec{p}_i)
= - [f^F(\varepsilon _i )]'  \,  p_i \,  V  \cos \theta _i $
 [compare such~$ \delta f_i$ with ansatz~(\ref{anz})].
Equation (\ref{relat}) corresponds to the following eigenvector
of the matrix~$\alpha^{(1)}$~(\ref{alf_res}):
\begin{eqnarray}
   \vec \Phi^{(1), V}  = \begin{pmatrix}
        1\\a
    \end{pmatrix}
\end{eqnarray}
with a zero eigenvalue, $\lambda^{(1),V}=0$:
\begin{eqnarray}
     \alpha^{(1)} \, \vec \Phi^{(1), V }=0
     \,.
\end{eqnarray}
We remind here that, as~$\beta^{(1)} =0 $, than the total relaxation
matrix is:~$ \Gamma^{(1)} =\alpha^{(1)} $.

Second, the following relation between the components of the wave function
with zero total particle flow, $j= j_1+j_2 =0$, appears:
\begin{eqnarray}
    \Phi^{(1)}_1(p_1^F)^3=-\Phi^{(1)}_2(p_2^F)^3
    \,.
    \label{relat2}
\end{eqnarray}
This relation corresponds to the expression for the flows~$j_i$:
 $ \vec{j}_i  = \sum _{\vec{p}_i} \vec{v}_i  \, \delta f _i (\vec{p}_i) $.
Relation~(\ref{relat2}) leads to the second eigenvector
of the matrix~(\ref{alf_res}):
\begin{equation}
   \vec \Phi^{(1), \, j}
   =\begin{pmatrix}
        1\\-a^3
\end{pmatrix}
\end{equation}
with a non-zero eigenvalue $\lambda^{(1) }$:
\begin{eqnarray}
\alpha^{(1)}  \,  \vec\Phi^{(1), \, j}
=
\lambda^{(1) }  \, \vec\Phi^{(1),\,j}
\:,
 \\
 \nonumber \\
\lambda^{(1)}={\alpha_{11}^{(1)}} \, (1+a^2)
\:.
\label{lll}
\end{eqnarray}

We also note that these eigenvectors~$\vec \Phi^{(1), V }$ and~$\vec \Phi^{(1), \, j}$
of the relaxation matrix of the first harmonics~$\alpha ^{(1)}$
can be obtained from qualitative reasoning. The first of them, with zero eigenvalue,
 corresponds to a state with zero relative velocity,
 that is, the state obtained from the equilibrium by the Galilean transformation.
 The second eigenvector describes a state with zero total current.
The momentum conservation law  for our system with the quadratic
energy spectrums of electrons in the two subbands, that leads to $\vec{v}_i = \vec{p}_i/m$,
 implies that the total current conserves, that is,
is independent of time. Correspondingly,
 the state with zero total current, $\vec j_1=-\vec j_2$, do relax due to
  the inter-subband scattering and, thus,
forms an eigenvector of the relaxation matrix~$\alpha^{(1)} = \Gamma ^{(1)} $.
Its eigenvalue~$\lambda^{(1)} $  is the relaxation rate
 of the relative velocity~$V=V_1-V_2$ of the fluid components.

It follows from the definition~(\ref{gammadefinition}) of $\Gamma ^{(1)}  = \alpha^{(1)} $
and the form of the distribution functions~$f_i$ corresponding to non-zero
flows $j_i = n_i V_i$,  that  the relaxation rates of the fluxes of each of the components
  are given by the formula:
\begin{equation}
\frac{d \vec j_\sigma }{dt}
=
-\sum _{\sigma ' } \:(S\alpha ^{(1)} S^{-1})_{ \sigma \sigma' }
 \: \vec j_{ \sigma' } \:,
\end{equation}
where the indexes $\sigma ,\sigma'$ refer to the first and second
spin subbands~``1'' and~``2'' and
\begin{eqnarray}
 \label{S}
S =
\begin{pmatrix}
        ( \, p_1^F \,)^3 &0
        \\\\
        0 & (\, p_2^F\,)^3
    \end{pmatrix}
\sim
\begin{pmatrix}
        a^3&0\\
        0&1
    \end{pmatrix}
\end{eqnarray}
is the matrix connecting the components of the functions $\Phi^{(1)}$
 and flows $\vec{j}_{1,2}$. From Eqs.~(\ref{alf_res}) and~(\ref{S})we get:
\begin{eqnarray}
\frac{d}{dt}\begin{pmatrix}
    \vec j_1\\\vec j_2
\end{pmatrix}
=
-\alpha_{11}^{(1)}\begin{pmatrix}
        1&-a^2\\-1&a^2
    \end{pmatrix}\begin{pmatrix}
        \vec j_1\\\vec j_2
    \end{pmatrix}\label{qwerty}
    \,.
\end{eqnarray}
We emphasize that the right-hand side of equation~(\ref{qwerty})
is proportional to the difference between  the hydrodynamic velocities
of the two components of the fluid, $\vec{V}_1- \vec{V}_2$, since
$\vec{j}_1 - a^2\vec{j}_2 = n_1(\vec{V}_1-\vec{V}_2) $.

Next, below we write down explicit formulas for the relaxation rates~$\alpha^{(2)}$
and~$\beta^{(2)}$   of the second harmonic of the distribution functions.

The diagonal element of the matrix $\alpha^{(2)}$, describing the contribution
to the rates from the inter-subband scattering, takes the form:
\begin{eqnarray}
 \label{alpha_11__2}
    \alpha_{11}^{(2)}
    =\frac{H}{p^F_1 p^F_2  }\int\limits_0^{2\pi}
    \frac{1-\cos2\hat\theta_3}{|\sin\theta_2|}|M'|^2d\theta_2
    = \nonumber \\
    = 4 C \int\limits_{-1}^1
    dx \: \frac{(a+x)^2}{a^2+2ax+1}
    \times \nonumber \\ \times
    \frac{ 1}{ \displaystyle \Big(
     \frac{ \sqrt{1-x^2} }{ \sqrt{2} r_s}+ \sqrt{a^2+2ax+1}
    \Big)^2}
    \:,
\end{eqnarray}
where constant $C$ is given by the same formula~(\ref{c}).

The result for $\alpha_{11}^{(2)}$ follows from formula~(\ref{alpha_11__2})
 after the change~$a =p_1^F/p_2^F  \to 1/ a$ and~$r_s = r_{s,1} \to r_{s,2} $.

The off-diagonal element of the matrix $\alpha^{(2)}$ turns out to be
zero according to formulas:
\begin{eqnarray}
 \label{alpha_12_2}
    {\alpha_{12}^{(2)} }=\frac{H p_2^F}{(p_1^F)^3}\int\limits_0^{2\pi}
    \frac{\cos 2\theta_2-\cos(2\hat\theta_3-2\theta_2)}{|\sin\theta_2|}|M'|^2d\theta_2
    \nonumber
    \\
=-4aD    \int\limits_{-1}^1
    \frac{dx(a+x)(1+ax)(a^2+2ax+1)^{-1}}
    {\left( \displaystyle \frac{ \sqrt{1-x^2} }{\sqrt{2} r_s}+2\sqrt{a^2+2ax+1}\right)^2}
    ,
\end{eqnarray}
leading to~$ \alpha_{12}^{(2)} = 0$.
Here the constant $D$ has the form:
\begin{eqnarray}
    D=
    4 \pi ^2 \Big( \frac{e^2 a_B^2}{\kappa}\Big)^2
    \frac{ H \, p_2^F}{  (p_1^F)^3} =
    \frac{\pi}{3} \frac{ m\, T^2 \, p_2^F}{  \hbar \, (p_1^F)^3}
    \:.
\end{eqnarray}
Indeed, all the integrals of the form [being a generalization of the integral
in Eq.~(\ref{alpha_12_2})]:
\begin{equation}
\label{I}
  I= \int\limits_{-1}^1
  F\left[\frac{1-x^2}{a^2+2ax+1}\right]
  \frac{(a+x)(1+ax)}{(a^2+2ax+1)^2}
  \, \textrm{d} x
\end{equation}
at $ a\neq\pm 1$ are equal to zero.
 To verify this, we will introduce a new parameter: $s=(a+1/a)/2$.
For $|a|\neq 1$,  we have $|s|>1$. Using these parameter,
we transform integral~(\ref{I}) as follows:
\begin{eqnarray}
 I=\phi(a)\int\limits_{-1}^1
  \tilde{F} \left[\frac{1-x^2}{x+s}\right]
  \frac{x^2+2sx+1}{(x+s)^2}\textrm {d} x
=
\nonumber\\=\phi(a)\int\limits_{-1}^1
  \frac{\textrm {d}~}{\textrm {d} x} \, G\left[\frac{1-x^2}{x+s}\right]
  \textrm {d}x=0
         \label{int}
         \:,
\end{eqnarray}
where the functions $\phi(a)$ and $\tilde{F}(z)$  directly come
from the function~$F(z)$ in Eq.~(\ref{I})
 via the change of variable,~$a \to s$, and
the function $G$ is the antiderivative of~$\tilde{F}(z)$:
$dG(z) / dz = \tilde{F}(z)$. The integral in Eq.~(\ref{int}) is zero,
since for $|s|>1$ the argument of the function $G$ is the same, zero,
 at the points $x=\pm 1$.

This just obtained result is rather non-trivial:  there is no effect
of second harmonic perturbations in the distribution function, $\Phi^{(2)}$,
 in one subband on the relaxation
of the second harmonic in the distribution function
 in the other subband: $\alpha_{12}^{(2)}=0,~~\ \alpha_{21}^{(2)}=0$.
 This fact can be qualitatively explained as follows.

\begin{figure}[t!]
 \centering
\centerline{\includegraphics[width=0.82 \linewidth]{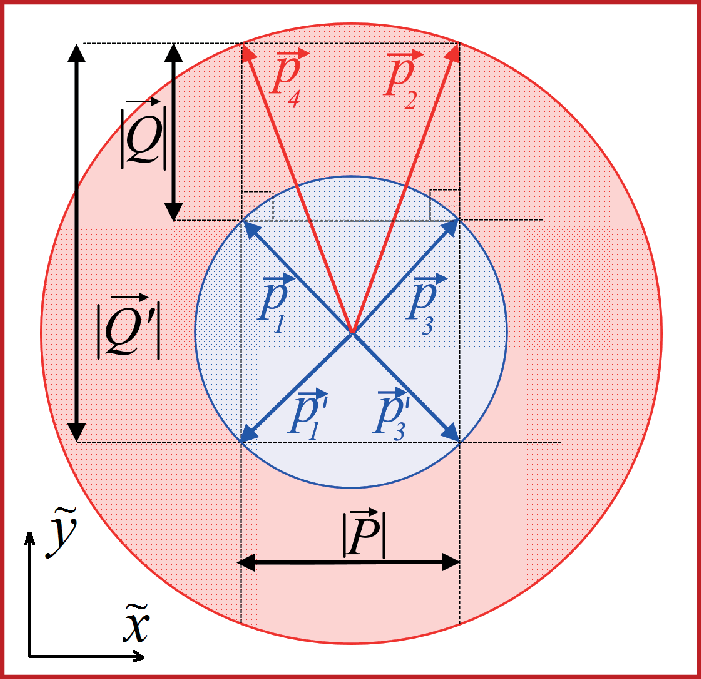}}
    \caption{ Scheme of kinematic relations between the momenta
    of the electrons in the inter-subband collisions.
    This scheme provides a visual explanation of
    the reasons for the absence of the ``second harmonic drag effect''.
     The two electrons with initial momenta~$\vec{p}_1$ and~$\vec{p}_2$
     or~$\vec{p}_1 \, '$ and~$\vec{p}_2$
     scatter to the states with the final momenta
     ~$\vec{p}_3$ and~$\vec{p}_4$      or~$\vec{p}_3^{\, '}$ and~$\vec{p}_4$.
      The initial distribution function in the red subband carries to a nonzero momentum flux,
      while the initial distribution function in the blue subband is equilibrium,
      without a momentum flux.
     The scattering  processes $12 \to 34 $ and~$1' \, 2 \to 3'\, 4 $,
     which are leading the the occurrence of
    inequilibrium perturbations of the distribution function in the blue subband
    due to the transfer of momenta in scattering,
     do not actually lead to a nonzero momentum flux in the blue subband
    due to the symmetry of the set
    of momenta~$\vec{p}_1 $, $\vec{p}_1 ^{\, '}$, $\vec{p}_3$, $\vec{p}_3 ^{\, '}$.
    All shown momenta lie near the Fermi surfaces, at the distances of the order
    of $T/v^F_i \ll p^F_i $ from it.
     The direction~$\tilde{y}$     corresponds to  the initial total momentum
    of the electron pairs,~$\vec{p}_1+\vec{p}_2$ and~$\vec{p}_{1} ^{\, '} +\vec{p}_2$,
    that is~$\vec{p}_{1,\, \widetilde{x}} +      \vec{p}_{2,\, \widetilde{x}} =0 $
    and~$\vec{p}_{ 1,\, \widetilde{x}} ^{\,'} +      \vec{p}_{ 2,\, \widetilde{x}} =0 $.
      Note that, if the scattering probability depended on the exchange vector~$Q$,
       the probability of scattering for the pairs,~$\vec{p}_1, \,\vec{p}_2$
     and~$\vec{p}_{1} ^{\, '} , \, \vec{p}_2$ would be different and
     the absence of the ``second harmonic drag effect''
      would  not be realized.
    }
    \label{nopi}
\end{figure}

Let us consider the transfer  process:
\begin{equation}
 \hat \Pi  _2 \sim \Phi _2^{(2)}
\to \hat \Pi  _1  \sim\Phi _1^{(2)}
\end{equation}
of  the momentum flux   due to the inter-subband scattering
from the red subband~``2'',  being initially  in a non-equilibrium state
carrying some non-zero momentum flux,   into the blue subsystem~``1'',
being initially in an  equilibrium state (see Fig.~3).
The matrix element $M'$ depends  on the transmitted
  momentum~$ \vec P=\vec p_1-\vec p_3 = \vec p_4-\vec p_2$,   and
do not depend on the exchange momenta~$\vec{Q}$  and~$\vec{Q}\,'$ (see Fig.~3),
 as we discussed in the above subsection.
The two, initial and final, momenta of electrons in the blue component~``1'',
involved in a collision, can have two different values: $\vec p_1\,, \; \vec p_3$
or~$\vec p_1\, ' \, ,\; \vec p_3\, ' $ (see Fig.~\ref{nopi}).
The scattering events:
\begin{equation}
  1\, 2 \to 3 \, 4 \: , \quad 1' \, 2 \to 3 '\, \, 4
  \:,
\end{equation}
 have the equal matrix elements,
both depending on $\vec{P}$, and, thus,  the equal probabilities.
Electron momentum flux tensor in the blue subband, which appears after
 these collisions  $1\, 2 \to 3 \, 4 $ and $1' \, 2 \to 3\, \, 4 $,
 is proportional to the following combination of a direct products
of the momenta~$\vec{p}_1$, $\vec{p}_1\,'$, $\vec{p}_3$,
 and~$\vec{p}_3\, '$:
\begin{eqnarray}
    \delta \hat{\Pi} \, \propto
\nonumber \\
 \propto
- \vec p_1 \otimes \vec p_1 
 \vec p_3\otimes\vec p_3-\vec p_1 \, ' \otimes\vec p_1 \, '
    + \vec p_3 \, ' \otimes \vec p_3 \, '
      \approx
    \nonumber
    \\
    \approx \,
    -\vec p_1\otimes\vec p_1 +\nonumber  \vec p_3\otimes\vec p_3 -
    \\
    -(-\vec p_3)\otimes(-\vec p_3) +(-\vec p_1)\otimes(-\vec p_1)
    =  0
    ,
    \label{tr}
\end{eqnarray}
here the sign~``$\approx$'' means the equality with a relative precision of
the order of the small parameter~$T / \varepsilon_F$.

From relation~(\ref{tr}) and~Fig.~3,  it is seen that
 the final momentum in the blue subband, which appears after the described collisions
 due to the presence of inequilibrium electrons in the red subband
(in the state $\vec{p}_2$ in Fig.~3)
 is relatively small as compared to the non-equilibrium momentum in the red subband,
to the extent of the small parameter~$T / \varepsilon_F$. So  is relatively weak
 the scattering-induced process of  change of the momentum flux in the blue subband
due to the presence the inequilibrium momentum flux  in the red subband.
The above reasoning can be easily generalized to the process of ``transferring''
of  the higher even harmonics.

Expressions for the intra-subband relaxation rates of the second angular harmonics,
 following from Eqs.~(\ref{res_beta_m_11}) and~(\ref{res_beta_m_22}),
take the form:
\begin{eqnarray}
 \label{beta_ii__2}
   { \beta_{ii}^{(2)}}
  = E_i
   \int\limits_0^{2\pi} d\theta _3  \: |\sin\theta _3 |
      \times
      \nonumber
         \\
    \times
  \Bigg[
   \frac{1}{ \displaystyle \frac{| \sin (\theta_3 /2) | }{\sqrt{2} r_{s,i}}+1}
    -
   \frac{1}{\displaystyle  \frac{| \cos (\theta_3 /2) |}{\sqrt{2} r_{s,i}} +1}
  \Bigg]^2
 ,
\end{eqnarray}
where  $r_{s,i} = 1/(a_B \sqrt{\pi n_i })$ and
\begin{equation}
E_i = 8 \pi ^2 \Big( \frac{e^2 a_B }{\kappa}\Big) ^2 \frac{H}{(p^F_i)^2}
 = \frac{2\pi}{3} \frac{ m\, T^2}{ \hbar \, (p^F_i)^2}
\,.
\end{equation}
Up to the form of the dimensionless matric element,
these expressions are identical with the expressions
obtained in Ref.~\cite{alekseev2020viscosity}
 for the one-component 2D electron system, degenerated by spin.

\section{ 5.  Results of calculation of relaxation rates  }
In this section, we describe the properties of relaxation rates
calculations using the formulas obtained above.

Figure 4 shows the calculated above relaxation rates
in the two-component 2D Zeeman electron system  formed
 from the quantum confinement level
 in a GaAs/AlGaAs quantum well in a strong tilted magnetic field.
The horizontal axis represents the Zeeman splitting
in units of the Fermi energy $\varepsilon_0^F$ of
the system without  magnetic field. Positive splitting values
correspond to the subband~``1''
(the red  subband with the smaller Fermi momentum),
while negative values correspond to the subband``1''
 (the blue  with the larger Fermi momentum).
 The graphs were plotted for realistic two-dimensional
 electron densities~$n$
 corresponding to a moderate interaction parameter,
 $r_s \sim 1$ [here we imply that $r_s=r_{s,0}=1/
(\sqrt{ \pi n}a_B)$], which lies on the border of applicability of our theory, as
the screened Coulomb potential~(\ref{U}) is valid only at~$r_s \ll 1 $.

It is seen from Fig.~4($a,c$) that the relaxation rates of the first and second
harmonics due to the inter-subband collisions are comparable.
We remind that that the relaxation rate of the first harmonic of the distribution function
due to intra-subband collisions is zero, $\beta_{ii}^{(2)} = 0$ (this follows simply from
 the momentum  conservation law within each subband).

One can see a logarithmic singularity in the dependencies
 $ \alpha _{11} ^{(1)} (B) $ and $ \lambda ^ {(1)} (B ) $ of the relaxation rates
 of the first angular harmonics (namely, the value~$\vec V_1- \vec V_2$,
 as it was shown
in the previous section) near zero magnetic field, that is at zero Zeeman
splitting when the two-component system becomes close to a one-component system
[see Fig.~4($a$,$b$)]. Mathematically, this singularity    is related to the fact that,
for example, the integral in Eq.~(\ref{alf_res__11})    at~$ a \to 1$
(corresponding to $B  = 0$) diverges logarithmically in  the lower limit~$x=-1$.
 In the limit $ a \to 1$ (at a fixed value of~$r_s =r_{s,0} $)
we have  from Eq.~(\ref{alf_res__11}):
\begin{equation}
  \label{div}
     \alpha _{11} ^{(1)} \, \sim \, C  \, r_s^2 \, \ln\Big(  \, \frac{1}{   |a-1|}\, \Big)\,.
\end{equation}
Note that $C= C(T) \sim T^2 /\varepsilon ^F_0$ at $a \to 1 $ [see Eq.~(\ref{c})].
This singularity  occurs as we do not take into account
the energy transfer during collisions between
electrons from different components. Indeed, in a one-component system
the head-on collisions predominate,  in which the energy transfer process
is critically important. Taking the energy transfer during collisions
into account  at small~$B$, when~$\mu B \sim T$, should lead to the elimination of divergence
 and smoothing of the peak. As a result, apparently, equation~(\ref{div}) will be replaced by:
\begin{equation}
   \alpha _{11} ^{(1) }
   \, \sim \,C \, r_s^2 \,
   \ln \Big( \,  \frac{ \varepsilon ^F _0  }{T } \, \Big)
   \,,
\end{equation}
where~$\varepsilon^F_0$ is the Fermi energy at zero magnetic field.
The last estimate is obtained from~(\ref{div}) just by replacing the difference  $|a - 1|$,
 characterizing the proximity of the the Fermi energies~$ \varepsilon ^ F_1 $
and~$\varepsilon ^ F_2$ one to another,  by its minimum possible value,
corresponding  to~$ | \varepsilon ^F_1 - \varepsilon ^F_2 | \sim T$.

The relaxation rate of the second harmonic due to
 the intra-band collisions,~$\beta_{ii}^{(2)}$, is relatively small
 as compared to the inter-subband rates,~$\alpha_{ii}^{(2)}$,  [compare Figs.~4($c$) and~4($d$)].
This originates from the fact that all particles within a single subband have the same spin,
 thus the space component of the wave function of two  colliding electrons
is antisymmetric [see Eq.~(\ref{intra-states})], leading to
a strong suppression of scattering processes
within one subband (a small probability to be close one to another due to the Pauli principle).
In other words, the latter relaxation rates,~$\beta_{ii}^{(2)}$,
 are non-zero only due to the non-point-like type of the interaction
 via  the screened Coulomb potential.

\begin{figure}[t!]
 \centering
\centerline{\includegraphics[width=1.02 \linewidth]{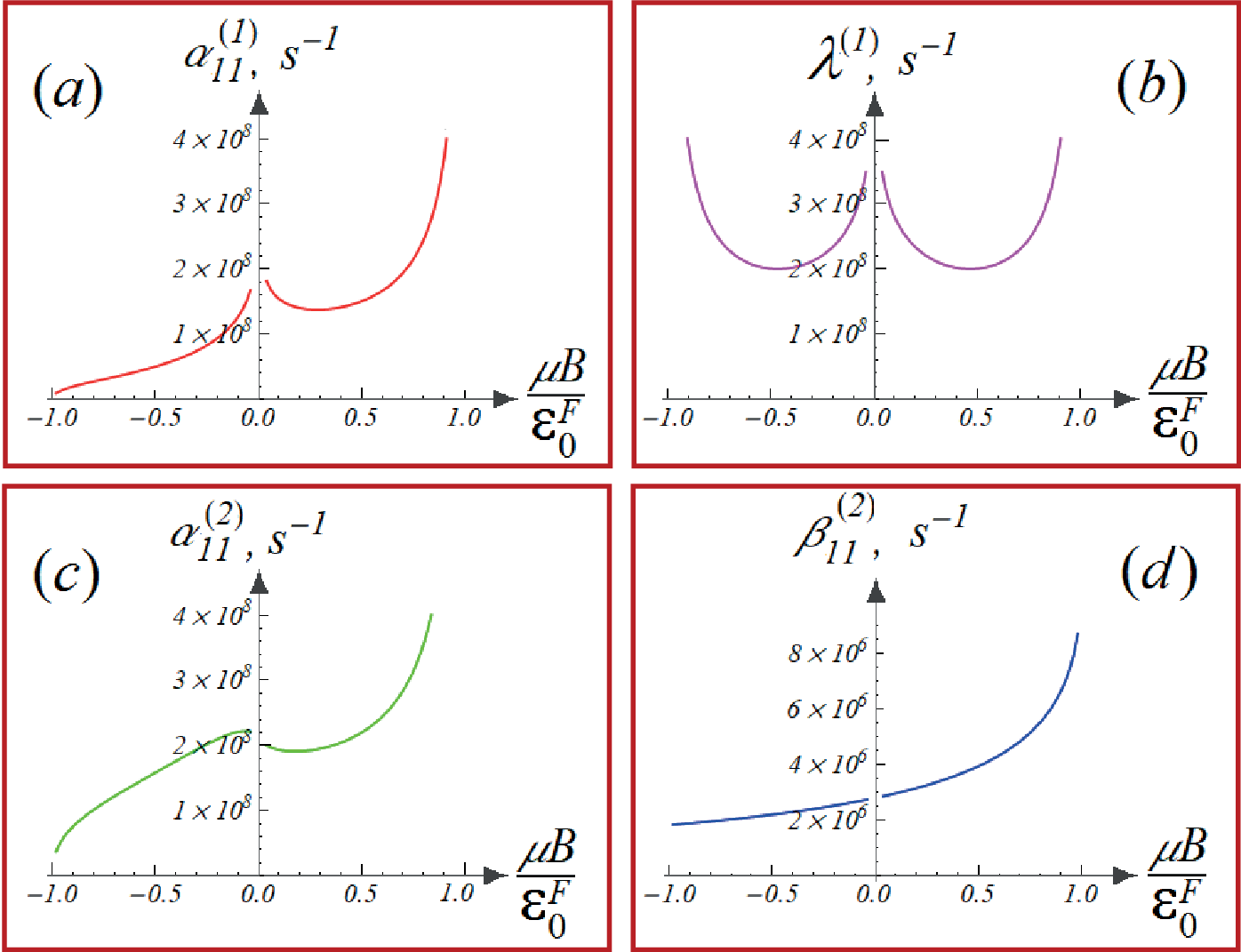}}
\caption{Relaxation rates
of the first and second angular harmonics of the distribution function,
calculated for the two-component electron system formed in quantum wells
in an inclined magnetic field (with the parameters corresponding to the sample
in experiment~\cite{dai2011response}). The horizontal axes present the values of
the Zeeman splitting~$\mu B \approx \mu B  $
in the units of the Fermi energy~$\varepsilon^F_0$ at~$B=0$.
The positive values~$\mu B \approx \mu B  $ of correspond
to the case~$p_1^F<p_2^F$ [which is shown in Fig.~1($a$)],
  the negative ones correspond to the opposite case $p_1^F>p_2^F$.
($a$):~Element~${\alpha_{11}^{(1)}}$ of the relaxation rate matrix
 of the first angular harmonic  (at positive values of~$B$).
   Due to the mentioned above inversion of the position
   of the subband~``1'' and~``2'' at~$B>0$ and~$B<0$,
 the plotted value at negative values of~$B$  is equal to $\alpha_{22}^{(1)}$
  for usually defined two-component system with~$p_1^F<p_2^F$.  The non-diagonal elements
  ${\alpha_{ij}^{(1)}}$, $ij = 12 , 21$, are expressed
   via ${\alpha_{11}^{(1)}}$ by Eq.~(\ref{alf_res}).
($b$):~The only non-zero eigenvalue $\lambda^{(1)}$~(\ref{lll})
of the relaxation matrix~$\alpha^{(1)}$
 of the first harmonics. The values~${\alpha_{ij}^{(1)}}$
 and~$\lambda^{(1)}$ diverges at $B \to 0$,
because in this limit the system becomes close by its properties to a one-component system.
($c$):~Elements of the relaxation matrix of the second angular harmonic
 corresponding to inter-subband collisions~${\alpha_{11}^{(2)}}$
 (in the positive part of the horizontal axis~$B$)  and~$\alpha_{22}^{(2)}$
   (in  the negative part of the horizontal axis~$B$).
$(d)$:~Elements of the relaxation matrix of the second moment corresponding
  to intra-subband collisions,~${\beta_{11}^{(2)}}$
    and~${\beta_{22}^{(2)}}$ (at $B>0$ and~$B<0$, correspondingly).
  We remind that the last values are much smaller than inter-subband
  rates~$\alpha_{ii}^{(2)}$
because of the Pauli principle.
 }
\label{f4}
\end{figure}

The following parameters, corresponding to high-quality GaAs quantum wells
 examined in Ref.~\cite{dai2011response},  were used for calculation:
$m=0.067 m_0$, $\kappa=13 $, $|g|=0.44$ (the effective mass, the dielectric constant,
and the electron $g$-factor in  GaAs),
$T=0.3~K$ and $n=3\cdot10^{11}cm^{-1}$ (the typical temperature and electron density
of 2D electrons in experiment~\cite{dai2011response}).
We have checked  that the general behavior of the dependencies
of the relaxation rates on~$B_x$
(the shape of the dependencies on  magnetic field)
remains qualitatively similar
 when the parameters of the system  are changed.

For the values of the dimensionless magnetic field parameter,
$b=\mu B/ \varepsilon^F_0 \sim 1$,
not too close to zero and to unity,
and in the limit $r_s \ll1,~~r_s\ll|1\pm b|$, analytical expressions
for the relaxation rates can be obtained.
This case corresponds to a two-component system with the subbands having substantially
different electron densities, each of which is not too small compared to the other.
Cutting the logarithmic divergences in Eqs.~(\ref{alf_res__11}), (\ref{alpha_11__2}),
and~(\ref{beta_ii__2}), we find:
\begin{equation}
  \left.
  \begin{array}{r}
  \lambda^{(1)}
  \\ 
 \alpha_{11}^{(2)}
  \\ 
 \beta_{11}^{(2)}
  \end{array}
  \right\}
   =
  C\, r_s^2 \ln\Big(  \frac{1}{r_s} \Big)
  \left\{
   \begin{array}{l}
   \displaystyle (1-b^2 )^{  -3/2}
   \\ 
   \displaystyle (1-b)^{-1} (1-b^2 )^{-1/2}
   \\  
\displaystyle    (1-b)^{-2}
   \end{array}
  \right.
  .
\end{equation}
We remind that for typical GaAs/AlGaAs quantum well samples,
studied in experiment~\cite{dai2011response} and others,
the densities are such that $r_s\sim 1$, while the limit $r_s \ll  1$
is supposes when we use the screened potential~(\ref{U}),
 that is the 2D electron system in real samples  lie
on the border of applicability of the developed theory.

If we use a simplified model of a point interaction potential, which corresponds
to a momentum-independent matrix element,
\begin{eqnarray}
   U(P)\,= \, const \, ,
\end{eqnarray}
 then all the rates
can be calculated analytically.  The results are generally similar
to those obtained in our model with a realistic screened Coulomb potential
(see Fig.~\ref{f4}).

\section{ 6. Hydrodynamic equations }
In this section, we formulate the hydrodynamic balance equations
 in the linear approximation for the considered  two-component electron system.
 Taking into account the obtained form of the first-
and second-harmonic relaxation tensors~$\alpha^{(1)}$ and~$\beta^{(1)}$,
we have rigorously derived,  within the well-known procedure~(see, for example,
 Refs.~\cite{alekseev2016negative,alekseev2020viscosity}),
 the closed system of the hydrodynamic Navier-Stokes-like equations for the flows
of the two fluid components.
Herewith in these equations we account only the electric field, the magnetic field,
having both the in-plane~$B_{||}$ and the perpendicular~$B_{z}$ components,
as well as the shear viscosity and the inter-subband friction.

For arbitrary 2D  flow geometry, such hydrodynamic equations for the flow
densities of the two components~$\vec j_i  ( \vec r,t)= n_i \vec  V_i ( \vec r,t) $,
$i=1,2$,   take the form:
\begin{equation}
 \label{hydr_eq_res}
\begin{array}{r}
\displaystyle
\frac{  \partial \vec j_1}{ \partial t}
=
 \frac{en_1}{m}\vec E+\omega_c[\vec j_1\times\vec e_z]+\eta_{xx,1}\Delta\vec j_1
\, +
\\\\ \displaystyle
+\,
\eta_{xy,1}[\Delta \vec j_1\times\vec e_z]-\lambda^{(1)}\frac{n_1n_2}{n}\vec V
\,,
\\\\ \displaystyle
\frac{  \partial \vec j_2}{ \partial  t}
=
 \frac{en_2}{m}\vec E+\omega_c[\vec j_2\times\vec e_z]+\eta_{xx,2}\Delta\vec j_2
\,  +
\\\\ \displaystyle
+ \,
\eta_{xy,2}[\Delta \vec j_2\times\vec e_z]+\lambda^{(1)}\frac{n_1n_2}{n}\vec V
\,,
\displaystyle
\end{array}
\end{equation}
where $\Delta$ is the Laplace operator, $\omega_c  = eB_z/(mc)$~is
the electron cyclotron frequency,
$\vec V=\vec V_1-\vec V_2=\vec j_1/n_1-\vec j_2/n_2 $ is the difference
between the flow velocities in the first and second subbands,
 $\lambda^{(1)}$~is the eigenvalue~(\ref{lll})
  of the relaxation matrix of the first moment $\alpha^{(1)}$.
Each subband has its own shear viscosity coefficient $\eta_{xx,i}$
and Hall viscosity coefficient $\eta_{xy,i}$:
\begin{eqnarray}
 \label{etas}
\eta_{xx,i}=\frac{(v_i^F)^2 \, \tau_{i} ^{(2)} /4}
 {1+\left(2\omega_c\tau_i^{(2)}\right)^2}
\,,
\;\;\;\;
\eta_{xy,i}=2\omega_c\tau_{i} ^{(2)} \, \eta_{xx,i}
\,.
\end{eqnarray}
We recall that the matrix of relaxation rates of the second moment~$ \Gamma ^{(2)}
_{ij} = \alpha_{ij}^{(2)} +
\beta_{ij}^{(2)} $ is diagonal, therefore, for each component of the fluid,
the relaxation time of the second angular harmonics is determined just
through the diagonal components, including both
the contributions from the intra- and inter
collisions:
\begin{eqnarray}
 \tau_{i} ^{(2)} \, = \,  1  \, / \,
  ( \,  \alpha_{ii}^{(2)} +  \, \beta_{ii}^{(2)}  \, )
 \,.
\end{eqnarray}

Equations (\ref{hydr_eq_res}) do not take into account
neither the diffusion contributions to the electron flows from
 the appearance of spatially inhomogeneous electron densities~$\delta n_i (\vec r , t)$,
 nor the relaxation  of the electron momentum induced by  the scattering
 in the bulk of the sample by residual defects or/and by acoustic phonons.
 The first effect arises due to rare electron transitions between the components
 (related to  the spin-dependent contribution
 in the  electron-electron electron scattering
and/or   the  spin-dependent scattering by defects) and is described
   by the balance equations for perturbations
   of the electron densities,~$\delta n _ i$.
Such equations for~$\delta n _ i$  allow to express the diffusion contribution,
$ \sim \nabla \delta n _ i $ in the equations for the time derivatives
of the flows~$ \partial \vec  j_i  / \partial t $
 via the second space gradients of
the flows~$ \partial ^2\vec  j_i  / \partial x _m \partial x _n  $:
\begin{equation}
 - \nabla \delta n _ i
  \, \sim \,
  \partial ^2\vec  j_i  / \partial x _m \partial x _n
\:,
\end{equation}
that makes up the effects of sort of the bulk viscosity.
 This approach was used to study  magnetotransport
in  two-component electron-hole systems in Refs.~\cite{we1,we3,we6,we4}  and
 in  two-component electron-electron systems
in Refs.~\cite{alekseev2023viscous,other_we7};
 for the case of electron-hole transitions due to interaction with acoustic phonons
 such equations for~$\delta n _ i$ were derived in Ref.~\cite{we6} from
the kinetic equation.

The effect of momentum relaxation of electrons in the bulk is described
by simple terms of the form:
\begin{eqnarray}
  - \, \vec{j}_{1,2} \,  / \,  \tau_{def,\,i}^{(1)}
  \:,
\end{eqnarray}
in the right-hand side of Eqs.~(\ref{hydr_eq_res}).
It becomes especially important at sufficiently low temperatures,
 when inter-electron collisions become very weak. Both  these two
 effects can have a significant impact on the magnetotransport  properties
of the two-component electron fluid.

Let us discuss the physical problems where the derived equations~(\ref{hydr_eq_res})
 with the additional effects described just above can be used.

First,  hydrodynamic equations~(\ref{hydr_eq_res}) allow  to study
 stationary magnetotransport in a two-component electron system.
For example, they can be solved for the simplest case of
a long rectangular sample with rough edges (the Poiseuille-like flow).
The total current and the Hall electric field of such sample can be
calculated as a function of the perpendicular and
in-plane magnetic field component.

Second, the derived equations (\ref{hydr_eq_res}) can allow one
  to study low-frequency flows of the two-component  fluid,   with frequencies
much lower than the frequencies  of all the calculated above relaxation rates:
\begin{equation}
  \omega \ll \alpha_{ij}^{(m)} ,\, \beta_{ij}^{(m)}
  \:.
\end{equation}
In particular, one can study slow flows of a two-component electron fluid
in long samples, which are analogous to the Womersley flow
in a conventional viscous liquid in a long sample.

Specific calculations of magnetoresistance and the Hall effect
based on the derived hydrodynamic equations~(\ref{hydr_eq_res}) are a subject
of a subsequent work, right here we present only the following qualitative
reasoning.

In Ref.~\cite{alekseev2023viscous}, it was shown that for
a two-component electron systems in defectless samples,
in which, along with the shear viscosity of each component,
rare electron transitions between the two components are possible,
both the strong negative magnetoresistance (being typical of hydrodynamic flows
in single-component systems~\cite{alekseev2016negative}) and
 the positive saturating magnetoresistance arise.

It is an important result of the current work that the friction between
the components,  expressed by the terms:
\begin{equation}
 \label{fr}
 \mp \, \lambda^{(1)}  \, \frac{n_1n_2}{n} \,  (\vec{V}_1-\vec{V}_2)
 \:,
\end{equation}
in Eqs.~(\ref{hydr_eq_res}), leads to the equalization of the hydrodynamic velocities
of the fluid and, consequently, to the effective merging of the two components.

In this way, we have calculated the kinetic coefficient $\lambda = \lambda ^{(1)} n_1 n_2/ n $
in the friction  term~(\ref{fr}),   related only  to the inter-subband scattering,
and the shear viscosity coefficients $\eta_{xx/xy, i}$~(\ref{etas}),
  related both  to the inter-subband and to the intra-subband scattering. They
constitute the set of the kinetic coefficients
 in the considered two-components systems, related to the relaxation of
 the  first and second angular harmonics of the distribution function and
   controlling the flows in  the hydrodynamic regime.
Other, more complex, kinetic coefficients,   related to perturbations of particle densities
 as well as heat flows, are also possible; they describe the heat conductivity,
  the bulk viscosity, and  the thermoelectric effects.

 We expect the following effect of  the inter-component friction on magnetoresostance 
  (relative to the perpendicular magnetic field component~$B_z$).
With the increase of the magnitude~$\lambda $ of this effect~(\ref{fr})
from zero to a sufficiently large value,
 the transport properties of the electron fluid with nearly independent components
(studied in Ref.~\cite{alekseev2023viscous}) should gradually disappear,
 and the two-component fluid will become increasingly close in its properties
 to a single-component fluid (see, for example, Ref.~\cite{alekseev2016negative}).
Since in these two limiting cases the magnetoresistance can be saturating positive
and is strongly negative, correspondingly,
such transition of the a flow type with the increase of the  inter-component friction
will be accompanied, apparently,
by a gradual change of magnetoresistance from the first type to the second one.
 In the process of such transition,
other types of complex (possibly, non-monotonous) magnetoresistance may also appear
within the derived model~(\ref{hydr_eq_res}),
 if we additionally account the inter-subband transitions
 via the balance equations for~$\delta n _ i$.

\section{ 7. Conclusion }
We have developed a microscopic theory of magnetotransport
in two-component 2D electron system.
We calculated the kinetic coefficients  describing the relaxation of the first and
the second angular harmonic of the electron distribution functions, which are
responsible for the formation of hydrodynamic
flows of the  two-components of the viscous electronic fluid.
 The calculations are based on the kinetic equation, taking into account
Zeeman splitting and without taking into account  spin-flip transitions between
subbands. The derived equations  differ from the phenomenological equations
used in work~\cite{alekseev2023viscous} by the term of relaxation
of the difference of  the hydrodynamic velocities of the two fluid
components.

Our results lead, in particular, to the following two conclusions.
(i)~Although a homogeneous    total electric current  does not relax  in
Inter-electron scattering in the system with the quadratic 
energy spectra of the electrons in the two component   (due to the momentum conservation law),
  scattering of the electrons in the two different subbands
  leads to the equalization of the hydrodynamic velocities
 in the two subbands. (ii)~The second angular
 harmonics of the electron distribution functions in  the two subbands,
 being responsible for the effect of shear viscosity, relax independently of each other.
The latter result is a consequence of the absence of an exchange contribution
 in the inter-particle interaction of electrons from different Zeeman subbands.
   It leads to significant simplifications of the hydrodynamic equations.

The obtained balance  equations and relaxation rates open up
the possibility of a quantitative description of magnetotransport effects
in two-component systems, in particular, the effects observed in
experiments~\cite{dai2011response,Hatke_Zudov,levin2024bulk}.
    Solving the transport equations derived here should make it possible
  to explain  the magnetoresistance observed in experiments,
  as well as
  to find the  conditions for the formation of
  hydrodynamic flows in two-component systems in various structures.
\\
\\
 \indent  This work was financially supported by the Russian
Science Foundation (Grant No. 25-12-00093).


\begin{thebibliography}{}
\bibitem{f}
We emphasize that, despite
 the successes in  the theory of the hydrodynamic transport
 in explanation of many bright experimemnts,
  many other key
experimental results remain puzzling. In particular,
  it is an open question up to now
  why signatures of the hydrodynamic transport, first of all,
  the giant negative magnetoresistance, regime persist down to
extremely low temperatures, when electron-electron collisions
are almost absent.
\\
\bibitem{small}
When we describe in this work the properties of the two-component  Zeeman systems,
we everywhere below imply a strong Zeeman splitting leading
 to substantial differences of the Fermi momenta and other parameter
of the two spin  subbands. At the same time,
  the hydrodynamic transport effects of 2D electrons
accounting the spin degree of freedom at zero
 or  small Zeeman  splittings,
 but in the presence of  the spin-orbit coupling
or at injection of spin-polarized electrons from contacts,
 were studied in several theoretical work, see,
for example, Refs.~\cite{Glazov}-\cite{Zohrabyan}.
\\
\end{thebibliography}
\end{document}